# Quantized thermal conductance in metallic heterojunctions

Nico Mosso[1,§], Alyssa Prasmusinto[1,2,§], Andrea Gemma[1], Ute Drechsler[1], Lukas Novotny[2] and Bernd Gotsmann[1,*]

[1]IBM Research, Zurich, 8803 Rüschlikon, Switzerland
[2]Department of Information Technology and Electrical Engineering, ETH, 8093 Zurich, Switzerland

[§]Both authors contributed equally to this work
[*]bgo@zurich.ibm.com

December 9,2018

To develop next-generation electronics and high efficiency energy-harvesting devices, it is crucial to understand how charge and heat are transported at the nanoscale. Metallic atomic-size contacts are ideal systems to probe the quantum limits of transport. The thermal and electrical conductance of gold atomic contacts has been recently proven to be quantized at room temperature. However, a big experimental challenge in such measurements is represented by the fast breaking dynamics of metallic junctions at room temperature, which can exceed the typical response time of the thermal measurement. Here we use a break-junction setup that combines Scanning Tunneling Microscopy with suspended micro electro-mechanical systems with a gold-covered membrane and an integrated heater acting also as thermometer. By using other metals as tip materials, namely Pt, PtIr and W, we show heat transport measurements through single gold atomic contacts. The dependence of the thermal conductance is analysed as function of contact size and material used. We find that by using Pt and Pt-Ir tips we can maximize the mechanical stability and probability of forming single Au atomic contacts. We then show the quantization of the electrical and thermal conductance with the verification of the Wiedemann-Franz law at the atomic scale. We expect these findings to increase the flexibility of experimental techniques probing heat transport in metallic quantum point contacts and to enable the investigation of thermal properties of molecular junctions.

## Introduction

Investigating the thermal transport properties of nanoscale metallic contacts is of fundamental interests for the scaling of electrical interconnects. Metallic atomic-size contacts represent the ultimate size limit and have been used in the last decades [1] as ideal systems to probe electrical conductance quantization.  Meanwhile, they also served as electrodes to contact single organic molecules and study their charge transport properties [2].

Quantization effects occur when the size of the conductor is comparable with the wavelength $\lambda_F$ of the charge carriers. Metallic atomic contacts are quantum one dimensional ballistic systems as the typical transversal size of about few atoms is on the order of the Fermi wavelength of the metal ($\lambda_F \sim 0.5$ nm [1]) and the length (up to few nm in the case of atomic chains) is well below the electron mean free path (10-100 nm) at room temperature [1,3]. In this regime, charge transport is usually described within the Landauer-Büttiker formalism [4], connecting the electrical properties of the mesoscopic conductor with the quantum mechanical transmission and reflection probabilities of the electron wavefunctions.

The quantization of the electrical conductance in atomic junctions has been mostly studied with Break-Junction techniques, namely Scanning Tunneling Microscopy (STM-BJ) or Mechanically Controlled Break Junction (MCBJ) [2]. These techniques consist of repeatedly forming and breaking nanoscopic metallic contacts while measuring the electrical conductance. The breaking process is characterized by a stepwise decrease of the electrical conductance due to the quantization effects and atomic rearrangements [5]. The formation of single atom contacts prior to complete rupture features a characteristic conductance value that depends on the number of electronic channels available for conduction and their transmission probability [6]. For instance, single gold atom contacts have a conductance of 1 $G_0$, where $G_0 = 2e^2/h$ is the electrical conductance quantum, $e$ the electron charge and $h$ the Planck constant, corresponding to a single electron channel with full transmission and spin degeneracy.

The break junction approach was adapted to investigate heat dissipation [7,8], thermoelectric effects [9,10] and, more recently, thermal transport [11,12] in such atomic junctions. It was demonstrated that the heat dissipation and thermoelectric properties are directly linked to the energy dependence of the transmission function [7,13]. Moreover, heat transport in gold and platinum atomic contacts is dominated by electrons with negligible phonon contribution [14] in agreement with the law of Wiedemann-Franz [11,15], which states that

$$G_{th} = L_0 \, T \, G_{el}$$

where $G_{th}$ is the thermal conductance, $L_0$ = 2.44·10$^{-8}$ V$^2$/K$^2$ the Lorenz number, $T$ the average junction temperature and $G_{el}$ the electrical conductance.

In our recent experiment [11], we studied heat transport across gold atomic contacts by using electrochemically etched gold tips on suspended Micro Electro-Mechanical Systems (MEMS) with integrated microheaters in a custom-built STM setup. In this work, we investigate the influence of the tip material on the thermal transport properties of gold atomic contacts. Remarkably, we find that electrochemically etched tips of Pt and Pt-Ir can be used without further treatment to form Au atomic contacts on the MEMS sensors with enhanced mechanical stability, showing a thermal conductance in good agreement with the Wiedemann-Franz law.

For STM imaging, the preferred tip materials are Pt-Ir alloys and W due to their durability, mechanical stability and ease of fabrication by electrochemical etching [16,17]. Indeed, early STM-BJ experiments were performed by indenting the W or Pt-Ir tip into a softer metal surface like Au. After few indentation steps, the tip surface is covered with Au atoms and no difference is observed in the electrical conductance traces and histograms with respect to experiments with pure Au tips [18–20]. This phenomenon can be explained as adhesion induced wetting of the hard tip by the softer metal, forming an atomically thin layer over the tip apex [21]. However, to avoid uncertainties in the chemical composition of the metallic contact formed, the same material for the tip and the surface has been typically selected in break junction experiments. This is especially important when studying the electrical properties of organic molecules, which depend critically on the binding configuration and the electrode work function [22,23].

Heat transport measurements introduce requirements on the tip radius and on the atomic junction stability, which usually do not represent a concern in STM-BJ measurements. Manually cut tips for instance are not suitable because of their large apex radius and irregular shape that increases the probability of forming parasitic contact with adsorbates on the surface [11]. Nanometer-sized tips are therefore preferred, which can be easily obtained by electrochemical etching of metal wires. However, electrochemically etched gold tips are often mechanically fragile and can lead to unwanted plastic deformation during the break junction measurement, decreasing the yield in the results.

Another important aspect in heat transport measurements is the lifetime of the atomic contact. The thermal conductance measurement features an intrinsic time constant proportional to the heat capacitance and thermal resistance of the MEMS sensor, which is on the order of few tens of ms. This defines the minimum time required to measure the thermal conductance of a single atom, in contrast with the electrical case, where the measurement bandwidth is only limited by the external electronics.

## Methods and Results

To investigate the effects of different tip materials on the heat transport measurements, we fabricated Au, Pt, Pt-Ir (80-20%) and W tips by electrochemical etching, adapting standard recipes [16,24–26] and achieving tip diameters below 100 nm. A detailed description of the fabrication process for the different metals is provided in Appendix A and B. The thermal conductance measurements were then performed on MEMS sensors as described previously [11]. Briefly, the MEMS consists of a suspended silicon nitride membrane with an integrated platinum microheater in thermal equilibrium with a gold platform to form and break contacts with the tip, Figure 1a. It is designed to have a high thermal resistance (~$2.5 \times 10^7$ K/W) with an estimated lateral stiffness of about 100 N/m.

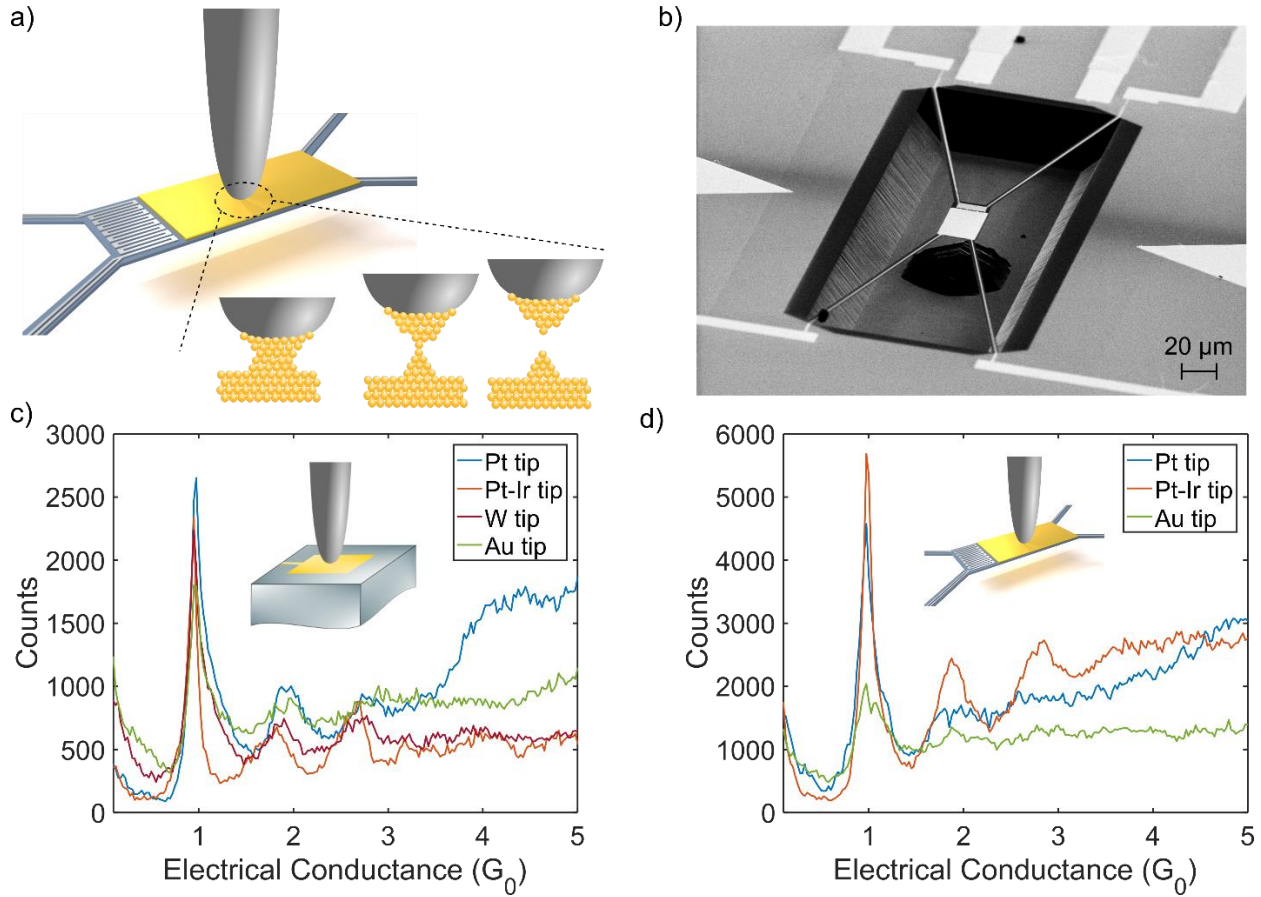

Figure 1. a) Schematic of the measurement setup showing the MEMS sensor in contact with the STM tip. b) SEM micrograph of the MEMS sensor. c) Electrical conductance histograms obtained with different tip materials on a non-suspended gold pad, built from 1000 opening traces using linear binning with a bin size of 0.025 $G_0$. d) Electrical histograms with tips made out of Pt, Pt-Ir and Au on the MEMS, built with respectively 1700, 2000 and 2000 traces using linear binning with a bin size of 0.025 $G_0$.

Before starting the measurement, the samples are cleaned with oxygen plasma (400 W , 5 min) to remove organic residues and then loaded into the vacuum chamber, waiting until the internal pressure reaches $10^{-7}$ mbar. During the experiment the tip and the sample substrate are thermally anchored to room temperature (22 °C) thanks to the mechanical links to the rest of the setup. The custom-built STM setup is located in the IBM Noise Free Labs, ensuring a temperature stability of 0.01 °C of the laboratory environment [27].

To test the STM tips, we first performed electrical STM-BJ measurements at a fixed small voltage below 50 mV on a non-suspended gold pad patterned on the substrate, which has undergone the same fabrication and cleaning steps as the gold platform on the MEMS. Note that the electrical current is measured with a series resistance of 13 kOhm and hence the actual voltage drop across the junction depends on the junction resistance itself. The results are summarized in Figure 1c. All the datasets were measured with a sampling time of 1 ms and pulling speed between 2 and 4 nm/s. For all the tips, it was possible to reproduce the typical histograms of Au-Au contacts, showing the characteristic peaks at multiples of the conductance quantum $G_0$ with similar intensities (number of counts). This is an indication for the formation of an Au metallic layer at the apex of the Pt, Pt-Ir and W tips. We note that the features in the conductance histograms can be used to identify the chemical species in the junction that are participating to the transport, as the d-wave metals like Pt and Ir, which were also found to form single atom contacts, do not show conductance quantization and the single atom conductance is between 1.5-2 $G_0$ for both metals [28–30]. Also in MCBJ experiments with metal alloys of Pt-Au and Pd-Au, the single atom peak at 1 $G_0$ appeared in the conductance histograms up to a gold concentration lower than 50% [31,32].

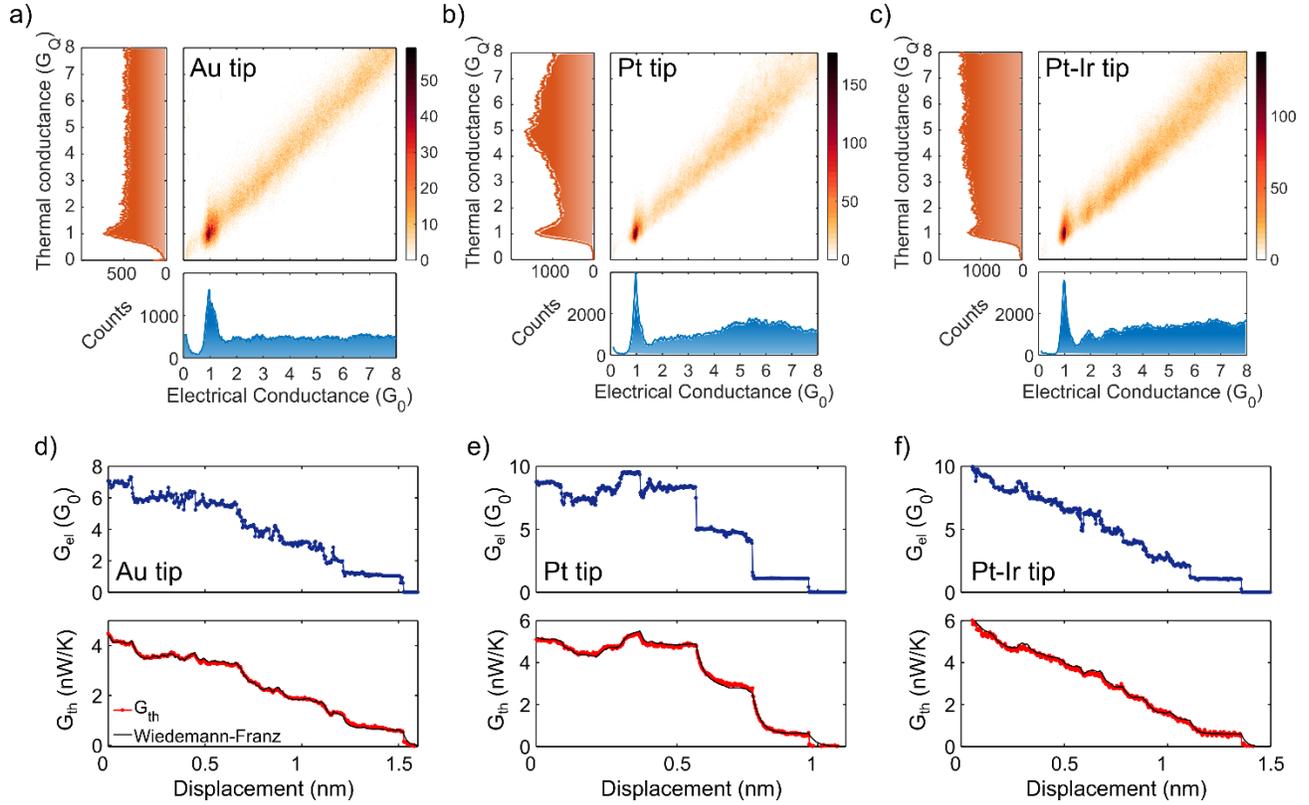

Figure 2 | a-b-c) 2D and 1D histograms of thermal versus electrical conductance for atomic Au contacts measured with Au, Pt and Pt-Ir tips, built with respectively 738, 877 and 982 traces, showing plateaus between 0.8 and 1.3 $G_0$ which are at least 10 ms in duration. d-e-f) Examples of single opening traces measured with different tip materials. The black continuous line in the thermal conductance plot is calculated by applying the Wiedemann-Franz law to the digitally low-pass filtered electrical trace.

Figure 1d shows the results obtained with similar tips on the MEMS, which was heated to about 60 °C to measure the thermal transport properties of the junction [11]. Measurements taken with W tips on the MEMS lead to inconclusive results, probably because of the fast oxidation of the tip surface combined with the MEMS mechanics. On the other hand, with Pt and Pt-Ir tips we observed pronounced peaks at 1 $G_0$ corresponding to the formation of single gold atomic contacts. Similar histograms were obtained if the tips were first used on the gold substrate or directly on the MEMS, indicating that the lower stiffness of the MEMS sensors did not influence the Au wetting process.

On average we can observe that the electrical conductance histograms measured with Pt and Pt-Ir tips show a greater number of counts at 1 $G_0$. This could result from a higher probability of forming single Au atom contacts or from an increased junction stability.

All the datasets were measured at a constant voltage of 40 mV, with a sampling time of 1 ms and a pulling speed between 2 and 4 nm/s along the motion direction of the tip.

Indeed, while on the non-suspended gold pad the tip is always approached perpendicular to the surface, on the MEMS the tip is typically moved at smaller angles to take advantage of the larger in-plane stiffness and improve the mechanical control over the breaking process, avoiding large jump out-of-contacts [11]. For these experiments, approaching angles between 20° and 30° were used, without noticeable differences.

We then measured simultaneously the electrical and thermal conductance of Au-Au contacts with the different tips. During the experiment the MEMS is heated to a temperature $T_H \sim 60$ °C by applying a constant voltage to the Pt-heater corresponding to few $\mu W$ of dissipated power. The temperature of the membrane is continuously monitored by measuring the 4-probe resistance of the Pt-heater and using the previously calibrated temperature coefficient of resistance α

$$R(T) = R_0(1 + \alpha \Delta T)$$

where $R_0$ is the heater 4P resistance at ambient temperature. The overall thermal conductance of the tip-MEMS system is calculated by dividing the total power dissipated in the heater and the temperature difference between the membrane and the tip/substrate $\Delta T = T_H - T_{amb}$.

The thermal conductance of the Au-Au contacts is then obtained by subtracting the out-of-contact value (tip in the tunneling regime), which corresponds to the contribution of the MEMS, and the in-contact one, which includes the contribution of the MEMS and the Au-Au junction.

Figure 2 d-e-f show typical thermal and electrical conductance opening traces measured with Au, Pt and Pt-Ir tips, respectively. Electrical and thermal conductance are linearly correlated, presenting steps and plateaus due to atomic rearrangements and conductance quantization at the same tip-MEMS displacements, and showing good agreement with the Wiedemann-Franz law. The delay observed in the thermal signal is due to the thermal time constant of the MEMS of about 20 ms.

Collecting several hundreds of such opening traces allows us to build 2 dimensional (2D) histograms to statistically investigate the relationship between thermal and electrical conductance, Figure 2 a-b-c. To reduce the low-pass filtering effects on the thermal signal, we focused the analysis on the traces with electrical plateaus lasting for at least 10 ms. Fast breaking curves, in fact, would only smear out the features in the thermal conductance histogram, hiding the single atom peak. For convenience, the thermal conductance was normalized by the degenerate thermal conductance quantum

$$G_Q = G_0 L_0 T = 2\frac{\pi^2 k_B^2 T}{3h}$$

where $k_B$ is Boltzmann constant, $h$ is Planck constant and $T$ is the average junction temperature. In this way, according to the Wiedemann-Franz law, all of the counts should accumulate around the diagonal of the 2D map. Noticeably, the thermal transport properties of the Au atomic contacts are independent of the tip material and in excellent agreement with the Wiedemann-Franz law, confirming that the phonon contribution to the thermal conductance is negligible within the experimental uncertainty (10%).

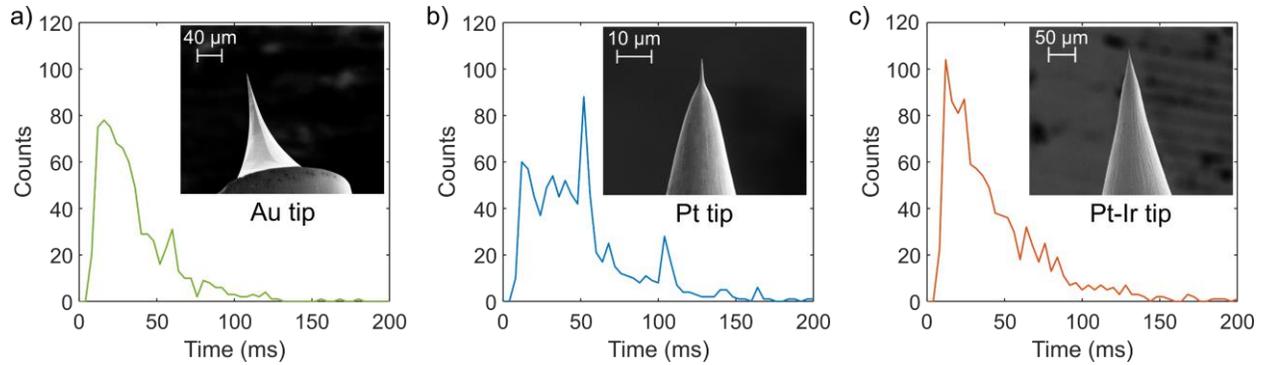

Figure 3. a-b-c) Histograms of the plateau duration in the electrical conductance range between 0.8 and 1.3 $G_0$ for the different tips. a) 738 traces. b) 877 traces. c) 982 traces. Very short plateaus (< 10 ms) were not included in the analysis. The insets show a scanning electron micrograph of the typical tip shape obtained with the electrochemical etching process.

Moreover, both the thermal and electrical 1D projections of the 2D histogram show a sharp peak centered at 1 conductance quantum.

This observation also implies that the interfacial thermal resistance of the gold layer adhering to the surface of the Pt and Pt-Ir tips is negligible with respect to the junction up to few atom contacts. Finally, at the actual junction between Au and Pt and Pt-Ir, respectively, Seebeck and Peltier effects may occur, the disregard of which could potentially lead to systematic errors in the analysis. However, we calculate that the thermoelectric effects caused by the metallic heterojunctions are negligible at these experimental conditions ($\Delta T$ = 40 K and V = 40 mV). This is also supported by the observation of the peaks in the thermal and electrical conductance at the conductance quantum.

## Discussion

The increased number of counts at 1 $G_0$ obtained with Pt and Pt-Ir tips may originate from a higher probability of forming single atom contacts or from an average longer plateau lifetime. To better understand this unexpected finding, we built histograms of the duration of the electrical 1 $G_0$ plateaus for the different tips, representing the lifetime distribution of the single Au atom contacts formed on the MEMS, Figure 3. Note that these histograms have been obtained from the same sets of traces shown in Figure 2. From the graphs, we can observe that the plateau distributions exhibit the same trend for the different tips suggesting a similar breaking mechanism. In particular, we can exclude the formation of single atomic chains with a good rate, as the peaks at multiples of the Au-Au interatomic spacing are missing in the histograms [33,34]. This is indeed expected at high temperature (~ 315 K) and at slow stretching rates where the breaking of the atomic contacts is mainly thermally activated [33].

The main difference is that for both Pt and Pt-Ir tips we find a probability of measuring 1 $G_0$ plateaus longer than 10 ms of about 50% compared to the 40% in the case of Au. Moreover, in the case of Pt, a large portion of traces shows an average duration of about 50 ms, suggesting the repeated formation of a stable junction configuration.

The mechanical stability of the Au atomic junctions may be influenced by several factors from the tip shape to the surface cleanliness. Understanding the microscopic details of these mechanisms requires further investigations that go beyond the scope of this work. However, by repeating the experiments with different tips and MEMS samples, we found that the results obtained with electrochemically etched Pt and Pt-Ir tips easily reproduce the best results obtained with Au tips. This indicates that having a mechanically stable tip made out of a stiff metal can be very beneficial in the case of heat transport measurements. Being flexible in terms of apex material may also help in the fabrication of tips with integrated thermocouples [12,35] and extend the use of Wollaston-like probes for such experiments.

## Conclusions

In summary, we showed that electrochemically etched Pt and Pt-Ir tips are well suited to study the heat transport properties of Au-Au atomic contacts. They are relatively easy to fabricate following standard recipes and can be used to form Au atomic contacts on Au surfaces with high probability and mechanical stability. This is extremely beneficial for thermal transport measurements which require a minimum measurement time of about one thermal time constant of the sensor used. We hope these results to foster the combination of STM imaging capabilities with thermal transport measurement at the atomic scale, which represents a fundamental stepping stone towards the control and understanding of heat transport mechanisms. Moreover, we expect that the increased experimental flexibility will enable to study the thermal conduction properties of single molecular electronics devices [36].

## Acknowledgements

The authors gratefully acknowledge management support from K. Moselund, W. Riess, H. Riel, M. Calame and J. Repp, and technical support from S. Karg, M. Tschudy, M. Sousa, S. Reidt, A. Olziersky, D. D. Pineda, F. Koenemann, and Y. Zemp.
We acknowledge funding by the European Commission H2020-FETOPEN 'EFINED' ( n⁰ 766853 ) and H2020-FETOPEN 'QuIET' (n⁰ 767187).

## Appendix A: Pt and Pt-Ir tips fabrication by electrochemical etching

To etch platinum and platinum-iridium tips, we developed a recipe starting from previously proposed methods [16,24]. Both groups suggested a two-step approach, comprising a coarse etching step using a relatively high AC voltage (> 25 V RMS) followed by a pulsed fine etching step at lower voltage (< 5 V). The first step is intended to form the main cone of the tip with a radius of a few micrometers, whereas the second forms a sharpened cone on top of the main one with an apex diameter of 100 nm, as shown in Figure 3b. The tips are made out of Pt (99.99+ %) and Pt-Ir (80%-20%) wire with 0.25 mm diameter (GoodFellow). A transformer connected to the power socket supplies an RMS voltage of 25 V at 50 Hz to the metal wire and the graphite (99.99+ %) counter electrode immersed in a 2M solution of $CaCl_2$, achieving sufficiently high current of around 500 mA. Etching times of about 4 min for platinum and 9 min for platinum-iridium are observed without drop-off. Then, a manual switch placed between the transformer and the counter electrode is used to apply voltage pulses of 2.5 V RMS in intervals of 1 s, achieving apexes with diameters of about 100 nm with a good reproducibility.

# Appendix B: W tips fabrication by electrochemical etching

Tungsten is another very common material used to fabricate sharp tips by electrochemical etching for scanning tunneling microscopy [17]. Among all the recipes available in literature, we adopted the method described by Collins [25], where a wire loop used as a counter electrode is not immersed in solution but contains a thin electrolyte layer, which the tungsten tip can go through. In this way, the etching is localized at the tip area in contact with the solution; the process ends with the drop of the portion below the loop, creating effectively 2 tips. Typically, the tip that falls shows the smallest radius. The technique works with both AC and DC voltages yielding a similar apex diameter of around 50-100 nm. The difference between them lies mainly in the etching time and the shape of the resulting tip cone: using AC modulation, the etching is fast and with bubbles (~ 3 min), forming a conical tip with a large cone angle, while using a DC voltage, it does not generate bubbles but it is rather slow (~ 12 min), giving tip shapes with high aspect ratio. For the STM-BJ experiments, the AC method is preferable as it provides tips with low aspect ratio, helping the formation of stable junctions.

After the etching, tungsten tips are known to oxidize quickly when exposed to air. There are several known methods to clean the oxide from STM tips, including UHV annealing, chemical treatment using hydrofluoric acid, AC polishing and sputtering [37]. We obtained the best results by cleaning the etched tips by ion milling (10 min, at 500 mA and 600 V). After this step, break junction experiments on a gold substrate yielded quantization peaks at multiples of the conductance quantum $G_0$, as expected for Au-Au contacts.

A gold loop is used as a back electrode while the W wire (99.95% - GoodFellow) with 0.25 mm diameter is immersed into a 2M NaOH solution. An AC voltage of 3.5 V at 50 Hz is then applied between the W wire and the gold loop, until the bottom part of the wire drops.